\documentclass[pra,aps,reprint,superscriptaddress,a4paper]{revtex4-1}
\usepackage{physics}
\usepackage{amsmath,amssymb}
\usepackage{braket}
\usepackage{graphicx}
\usepackage{xcolor}
\usepackage{verbatim}
\usepackage{color}
\usepackage[colorlinks, linkcolor={blue}, citecolor={blue}, urlcolor={blue}]{hyperref}
\usepackage{tabularx}
\usepackage{booktabs}
\usepackage{layouts}
\usepackage{blindtext}
\begin{document}
\title{Interacting bosonic flux ladders with a synthetic dimension:\\Ground-state phases and quantum quench dynamics}
\author{Maximilian Buser}
\affiliation{
Department of Physics,
Arnold Sommerfeld Center for Theoretical Physics (ASC),
Munich Center for Quantum Science and Technology (MCQST),
Fakult\"{a}t f\"{u}r Physik, Ludwig-Maximilians-Universit\"{a}t M\"{u}nchen,
80333 M\"{u}nchen, Germany
}
\author{Claudius Hubig}
\affiliation{
Max-Planck-Institut f\"ur Quantenoptik, Hans-Kopfermann-Strasse 1, 85748 Garching, Germany
}
\author{Ulrich Schollw\"ock}
\affiliation{
Department of Physics,
Arnold Sommerfeld Center for Theoretical Physics (ASC),
Munich Center for Quantum Science and Technology (MCQST),
Fakult\"{a}t f\"{u}r Physik, Ludwig-Maximilians-Universit\"{a}t M\"{u}nchen,
80333 M\"{u}nchen, Germany
}
\author{Leticia Tarruell}
\affiliation{
ICFO -- Institut de Ciencies Fotoniques, The Barcelona Institute of Science and Technology, 08860 Castelldefels (Barcelona), Spain
}
\author{Fabian Heidrich-Meisner}
\email{Corresponding author: heidrich-meisner@uni-goettingen.de}
\affiliation{
Institute for Theoretical Physics, Georg-August-Universit\"at G\"ottingen, 37077 G\"ottingen, Germany
}
\date{\today}
\begin{abstract}
Flux ladders constitute the minimal setup enabling a systematic understanding of the rich physics of interacting particles subjected simultaneously to strong magnetic fields and a lattice potential.
In this paper, the ground-state phase diagram of a flux-ladder model is mapped out using extensive density-matrix renormalization-group simulations.
The emphasis is put on parameters which can be experimentally realized exploiting the internal states of potassium atoms as a synthetic dimension.
The focus is on accessible observables such as the chiral current and the leg-population imbalance.
Considering a particle filling of one boson per rung, we report  the existence of a Mott-insulating Meissner phase as well as biased-ladder phases on top of superfluids and Mott insulators.
Furthermore, we demonstrate that quantum quenches from suitably chosen initial states can be used to probe the equilibrium properties in the transient dynamics.
Concretely, we consider the instantaneous turning on of hopping matrix elements along the rungs or legs in the synthetic flux-ladder model, with different initial particle distributions.
We show that clear signatures of the biased-ladder phase can be observed in the transient dynamics.
Moreover, the behavior of the chiral current in the transient dynamics is discussed.
The results presented in this work provide guidelines for future implementations of flux ladders in experimental setups exploiting a synthetic dimension.
\end{abstract}	
\maketitle
\section{Introduction}
The last decade has witnessed tremendous progress in the realization of artificial gauge fields in quantum engineered systems~\cite{cooper_19, aidelsburger_18, goldman_16, goldman_14, galitski_13, dalibard_11}.
In this context, magnetic fields have been emulated in different types of artificial lattice systems, including superconducting circuits~\cite{roushan_17, roushan_17b}, photonic setups~\cite{wang_09, fang_13, ningyuan_15, owens_18}, and ultracold quantum gases~\cite{aidelsburger_11, struck_12, aidelsburger_13, miyake_13, atala_14, jotzu_14, stuhl_15, mancini_15, livi_16, flaschner_16, an_17,tai_17,kolkowitz_17,kang_18, an_18, han_19, genkina_19, asteria_19, chalopin_20}.
In particular, for ultracold quantum gases, effective Peierls phases have been implemented by means of laser-assisted hopping~\cite{aidelsburger_11, aidelsburger_13, miyake_13} or Floquet engineering~\cite{struck_12, jotzu_14, flaschner_16, asteria_19}.
Moreover, the coherent coupling of the internal atomic states using optical transitions represents a very promising approach, addressing quasi-one-dimensional lattices.
This method  has been proposed in Ref.~\cite{celi_14} and successfully employed in several experiments~\cite{stuhl_15, mancini_15, livi_16, kolkowitz_17, han_19, genkina_19} for the emulation of charged particles in ribbonlike lattices pierced by uniform magnetic fields, dubbed flux ladders.
In this context, the atoms are subjected to a periodic real-space potential along the legs of the ladder, while a coherent coupling between the internal atomic states constitutes the rungs of the ladder, realizing a synthetic dimension.
Alternative schemes for the experimental realization of flux ladders include the use of other degrees of freedom to implement a synthetic dimension~\cite{an_17, kang_18, an_18}, or the isolation of ribbons in real-space two-dimensional Hofstadter systems using superlattice potentials~\cite{atala_14}, or digital micromirror devices~\cite{tai_17}.
The wide range of available atomic species~\cite{bloch_08}, the existence of Feshbach resonances~\cite{chin_10}, and the promising complementary approaches for implementing Abelian gauge fields pursued in the quantum-gas community render cold gases a promising platform for the realization of topology in interacting quantum matter~\cite{cooper_19, aidelsburger_18, goldman_16, goldman_14, galitski_13, dalibard_11}.
%

%
Flux ladders constitute a minimal setup allowing for the interplay between effective magnetic fields and interparticle interactions.
Theoretical studies have shown that due to this interplay, they host rich ground-state phase diagrams, including vortex-liquid and Meissner phases inherited from the weakly interacting regime~\cite{orignac_01, huegel_14}, which can exist on top of superfluids and Mott insulators~\cite{petrescu_13,piraud_15}.
Also, they feature ground states breaking a discrete symmetry, such as vortex-lattice~\cite{orignac_01, dhar_12,dhar_13,greschner_15, greschner_16}, charge-density-wave~\cite{greschner_16,greschner_17}, and biased-ladder states~\cite{wei_14}.
Moreover, the possible existence of Laughlin-like states has attracted great interest~\cite{grusdt_14, petrescu_15, cornfeld_15, petrescu_17, strinati_17, strinati_19b} and the study of the Hall effect in flux ladders remains an active line of research~\cite{prelovsek_99, zotos_00, greschner_19, filippone_19}.
The ground-state phase diagram of the two-leg flux-ladder model has been discussed in detail and mapped out to a large extent within previous theoretical studies~\cite{carr_06,roux_07,tokuno_14,didio_15,uchino_15,keles_15,uchino_16,bilitewski_16,orignac_16,orignac_17,huegel_17,citro_18,strinati_19}.
In many of these studies~\cite{roux_07,didio_15,keles_15,bilitewski_16,orignac_16,orignac_17,citro_18,strinati_19}, the density-matrix renormalization-group method~\cite{white_92, schollwoeck_05, schollwoeck_11} has been the numerical method of choice.
%

%
However, the exploration of the exact parameter regimes that could be accessed in future experiments, including the influence of nearest-neighbor rung-wise interactions, which are typically present in synthetic dimension implementations, remains an important open question.
The same applies to the investigation of the role of finite energy densities and temperatures~\cite{greschner_15,strinati_17,citro_18,coira_18,buser_19} on the ground-state phase diagrams~\cite{greschner_16}, and the development of optimal state-preparation protocols~\cite{wang_20} (based, for instance, on the dynamics induced by quantum quenches in the flux-ladder model).
At the same time, proposing adequate methods to probe and detect the different phases, i.e., via spectroscopic~\cite{wall_16, strinati_18, repellin_19}, transport~\cite{wang_13, zeng_15, mugel_17, taddia_17}, or microscopic measurements, remains an outstanding issue.
%

%
So far, experimental realizations of flux ladders have mostly concentrated on the non- or weakly interacting regime.
The strongly interacting regime has remained elusive due to the existence of detrimental heating processes associated to most of the experimental methods used to emulate strong magnetic fields~\cite{eckardt_17}.
A step towards exploring the many-body case was the experimental study of the dynamics of two repulsively interacting bosons on a real-space flux ladder~\cite{tai_17}.
Synthetic-dimension implementations seem particularly promising to extend these studies to a larger number of particles as we will discuss in detail in our work.
%

%
\begin{figure}
	\includegraphics[]{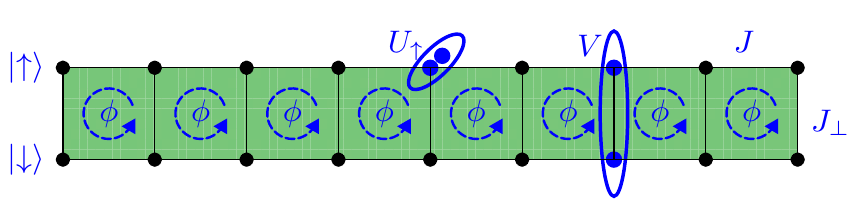}
	\caption{
		Two-leg flux-ladder model.
		In setups exploiting synthetic dimensions, the legs of the ladder correspond to different internal atomic states $\ket{\uparrow}$ and $\ket{\downarrow}$ as discussed in Sec.~\ref{sec:model:exp}.
		The Hamiltonian parameters $J$, $J_\perp$, $V$, $U_\uparrow$, and $\phi$ are introduced in Eq.~\eqref{hamiltonian}.}
	\label{ladder_sketch}
\end{figure}
In this paper, we focus on bosonic two-leg synthetic flux ladders implemented exploiting two particular internal states of $^{41}$K and on a value of the magnetic flux that is particularly simple to obtain experimentally.
The envisioned setup using potassium will first be described to motivate our choice of model parameters, particle filling, and initial states based on the experimental feasibility.
%

%
Second, we map out the ground-state phase diagram of the interacting flux-ladder model, which is illustrated in Fig.~\ref{ladder_sketch}, using extensive density-matrix renormalization-group simulations.
Putting the emphasis on a particle filling of one boson per rung and on parameters which are particularly well suited for the envisioned future experiments, we report on the existence of a biased-ladder phase~\cite{wei_14} for intermediate interaction strengths (see Ref.~\cite{greschner_15,greschner_16} for previous numerical studies of this phase).
Most notably, the biased-ladder phase exhibits a population imbalance between the legs of the system, which is typically stabilized by rung-wise interactions.
Additionally, we show that the system undergoes a superfluid-to-Mott insulator transition within the biased-ladder region for increasing interaction strengths, which is revealed by the opening of a charge gap.
In the studied parameter range, the biased-ladder phase neighbors a  Meissner phase~\cite{orignac_01, petrescu_13, piraud_15}. 
The Meissner state is typically found by increasing the rung hopping amplitude and, in the limit of vanishing interactions, corresponds to a single minimum in the one-particle dispersion.
The properties of the three ground-state phases considered in this paper are summarized in Table~\ref{tab:gsphases}.
\begin{table}
	\caption{\label{tab:gsphases}
		Overview over the three ground-state phases considered in this paper, listing characteristic values of the leg-population imbalance $\Delta_m$, the charge gap $\Delta\mu$, and the central charge $c$, as discussed in the main text.
		Note that various ground states in the two-leg flux ladder are comprehensively discussed in Ref.~\cite{greschner_16}, which also provides an extended tabular overview.
	}
	\begin{ruledtabular}
		\begin{tabular}{l c c c}
			& $\Delta_m$ & $\Delta\mu$ & $c$ \\ \hline 
			Biased-ladder phase~(superfluid) & $>0$ & $0$ & $1$\\
			Biased-ladder phase~(Mott-insulating) & $>0$ & $>0$ & $0$\\
			Meissner phase~(Mott-insulating) & $0$ & $>0$ & $0$\\
		\end{tabular}
	\end{ruledtabular}
\end{table}
Note that their existence in flux-ladder models has been discussed in previous works~\cite{orignac_01, petrescu_13, wei_14, piraud_15, greschner_16}.
However, hitherto, the regime of parameters relevant to ultracold potassium gases has remained unexplored.
In our analysis, we focus on accessible observables such as the chiral current and exemplify typical particle-current patterns and momentum-distribution functions in various ground-state phases.
%

%
Third, by time-evolving matrix-product states~\cite{haegeman_11, paeckel_19}, we investigate experimentally feasible quench protocols.
They allow us to probe characteristics of quantum phases that are otherwise hardly accessible due to the notorious difficulty of low-entropy state preparation at certain model parameters.
Noting that control over hopping matrix elements is well established in synthetic flux ladders~\cite{ stuhl_15, mancini_15, genkina_19}, we show that for the instantaneous turning on of leg hopping in the Meissner phase, chiral currents in the short-time dynamics exhibit a similar dependence on the model parameters as the corresponding ground-state currents.
Moreover, we show that clear signatures of the biased-ladder phase can be observed in the time evolution of the density imbalance between the legs, following the instantaneous turning on of rung hopping.
%

%
This paper is organized as follows.
In Sec.~\ref{sec:model}, we describe our proposed implementation of the flux-ladder model, define the Hamiltonian of the system and introduce several observables of interest.
Section~\ref{sec:numerical_methods} comments on the matrix-product-state methods used in this paper.
In Sec.~\ref{sec:gsphases}, we review properties of known ground-state phases in the flux-ladder model.
We map out the zero-temperature phase diagram in Sec.~\ref{sec:zeroT_PD}.
Quantum quenches in the synthetic flux-ladder model are discussed in Sec.~\ref{sec:quench_dyn}.
Finally, we summarize our work in Sec.~\ref{sec:summary}.
\section{Synthetic flux-ladder model}
\label{sec:model}
\subsection{Experimental scheme}
\label{sec:model:exp}
In this paper, we consider synthetic flux ladders, i.e.,  flux ladders realized using one real dimension and one synthetic dimension consisting of two internal atomic states, denoted in the following by $\uparrow$ and $\downarrow$.
Our proposal for an experimental implementation is the following.
We strongly confine the atoms in two transverse directions, in order to realize a one-dimensional Bose gas and subject them to a spin-independent one-dimensional optical lattice potential.
In this situation, each spin state realizes one leg of the ladder and nearest-neighbor hopping along the legs is determined by the leg hopping rate $J$, which can be tuned by the depth of the optical lattice.
In addition, we coherently couple the spins via two-photon Raman transitions, effectively implementing a rung hopping rate $J_{\perp}$ along a synthetic spin dimension.
The momentum transferred to the atoms by the Raman lasers during the spin flip generates Peierls phases for the motion along the rungs, mimicking the effect of a magnetic field piercing the ladder~\cite{celi_14}.
%

%
So far, a major limitation of  the experimental study of flux ladders with ultracold atoms has been the presence of detrimental heating processes associated to the periodic driving methods used to generate an artificial magnetic field~\cite{Messer2018,Viebahn2020,eckardt_17,reitter_17}.
Although synthetic flux ladders are also driven systems, the driving frequency corresponds to the Zeeman splitting between atomic sublevels and is several orders of magnitude larger than in Floquet-based schemes.
In this situation, one expects negligible heating caused by the coupling to the atomic micromotion.
Therefore, these systems constitute an appealing experimental platform for exploring the phase diagram of interacting bosons on flux ladders.
%

%
For concreteness, we focus  on the implementation of such flux ladders exploiting two internal states of $^{41}$K.
Furthermore, we consider the simplest experimental situation where both the optical lattice and the Raman coupling are produced by counter-propagating laser beams~\cite{stuhl_15}.
We set the wavelength of the Raman lasers to the potassium tune-out value $\lambda_R=769$ nm, which maximizes the coupling strength and does not produce any scalar potential on the atoms.
For an optical lattice created by a laser of wavelength $\lambda_L=1064$ nm, this yields a magnetic flux $\phi/(2\pi)=\lambda_L/\lambda_R=1064/769$, which will therefore be the value employed in the numerical simulations.
%

%
The strong transverse confinement required to enter the one-dimensional regime is realized with two additional optical lattices of wavelength $\lambda_L$, which propagate along the transverse directions.
We select a large depth in the range $V_L\sim 40\mbox{--}55\,E_L$, where $E_L=h^2/(2 m \lambda_L^2)$ is the lattice recoil energy, $h$ is Planck's constant, and $m$ is the mass of $^{41}$K.
This leads to the creation of an array of one-dimensional systems with negligible hopping between them during the timescales considered in this work.
%

%
As internal atomic states, we select the Zeeman sublevels $m_F=-1$ and $m_F=0$ of the $F=1$ hyperfine manifold of $^{41}$K, and use the notation $\ket{\uparrow}\equiv\ket{F=1,m_F=-1}$ and $\ket{\downarrow}\equiv\ket{F=1,m_F=0}$.
Our choice is motivated by the interaction properties of these states.
The $\uparrow\uparrow$ and $\downarrow\downarrow$ collisions are described by very similar and positive scattering lengths, resulting in nearly identical repulsive interactions in the legs of the ladder.
In contrast, interactions in the $\uparrow\downarrow$ channel can be controlled exploiting an interstate Feshbach resonance located at a magnetic field $\sim52$ G~\cite{lysebo_10, tanzi_18}.
In principle, it should be possible to vary the sign and strength of the interactions along the rungs, or even to completely cancel them.
Achieving the necessary magnetic field stability might, however, be challenging.
Therefore, in this paper, we focus on a large magnetic field limit ($\sim400$ G), where the two-photon Raman transitions are essentially immune against magnetic field fluctuations.
In this situation, small rung hopping rates $J_{\perp}<h\times100$ Hz should be within reach, making the regime of both small ($J_{\perp}/J<1$) and large ($J_{\perp}/J>1$) rung-to-leg hopping rate ratios experimentally accessible.
In the following, we consider the full range $J_{\perp}/J=0.2\mbox{--}30$.
In this magnetic field regime, the $^{41}$K scattering lengths are essentially identical ($a_{\uparrow\uparrow}=60.89\,a_0$, $a_{\downarrow\downarrow}=60.85\,a_0$ and $a_{\uparrow\downarrow}=60.72\,a_0$, where $a_0$ is the Bohr radius) and the system is nearly SU(2) symmetric.
This is, therefore,  the situation considered in the simulations.
Controlling the longitudinal optical lattice depth in the range $V_L\sim 4\mbox{--}10\,E_L$ allows one to adjust the interparticle-interactions-to-hopping ratio $U/J$, and to realize values $U/J=2.5\mbox{--}20$.
Note that the second bosonic isotope of potassium, $^{39}$K, should also allow to explore  situations where the interactions in each leg are very different, or even have opposite signs \cite{derrico_07, lysebo_10, cabrera_18, cheiney_18, sanz_19}.
Studying these configurations, which are expected to give rise to density-dependent Peierls phases, goes beyond the scope of this work.
%

%
Finally, in all simulations, we consider a filling of one particle per rung.
This situation could be easily obtained by starting with a system occupying a single leg of the ladder (i.e., a single spin state), in a Mott-insulator state with one particle per site, and with negligible leg hopping $J$.
Then, activating the rung hopping $J_{\perp}$ by turning on the coupling between the two spin states yields the desired filling of one particle per rung.
This preparation sequence also allows one to realize the initial states of the quench protocols discussed in Sec.~\ref{sec:quench_dyn} by adjusting the initial values of the leg and rung hoppings before the quench.
The experimental ingredients described here for this future experiment are therefore all readily available.
\subsection{Hamiltonian}
The Hamiltonian describing the synthetic flux-ladder model is
\begin{align}
H =
&- J\sum_{\sigma=\uparrow, \downarrow}\sum_{r=0}^{L-2} \left( a_{r,\sigma}^\dagger a_{r+1,\sigma}+\text{h.c.} \right)\nonumber\\
&-J_\perp\sum_{r=0}^{L-1}\left( e^{-ir\phi}  a_{r,\uparrow}^\dagger a_{r,\downarrow}+\text{h.c.} \right)\nonumber\\
&+\frac{1}{2}\sum_{\sigma=\uparrow, \downarrow}^{1}U_\sigma \sum_{r=0}^{L-1}  n_{r,\sigma}\left( n_{r,\sigma}-1\right)\nonumber\\
&+V\sum_{r=0}^{L-1}  n_{r,\uparrow} n_{r,\downarrow},
\label{hamiltonian}
\end{align}
with the parameters $J$ and $J_\perp$ corresponding to nearest-neighbor hopping along the legs and rungs of the ladder, respectively, as described in Sec.~\ref{sec:model:exp}.
The site-local operator $a^{(\dagger)}_{r,\sigma}$ annihilates (creates) a boson on site $(r,\sigma)$.
Note that the internal atomic states $\sigma=\uparrow,\downarrow$, also introduced in Sec.~\ref{sec:model:exp}, are identified with the legs of the ladder.
Further, $n_{r,\sigma} = a_{r,\sigma}^\dagger a_{r,\sigma}$ accounts for the occupation of local lattice sites.
It is worth noting that we consider the so-called rung gauge~\cite{greschner_16} in which the Peierls phase factors are aligned along the rungs of the ladder.
They are chosen in such a way that whenever a particle encircles a single plaquette of the ladder, its wavefunction gains a phase factor $e^{\pm i \phi}$, with the sign depending on the direction of the circulation.
The parameters $U_\uparrow$ and $U_\downarrow$ determine site-local interparticle interactions on the $\uparrow$ leg and $\downarrow$ leg, respectively, while $V$ accounts for rung-wise interactions.
In experimental implementations based on synthetic dimensions, the former are associated with the $a_{\uparrow\uparrow}$ and $a_{\downarrow\downarrow}$ scattering lengths, while the latter is proportional to the value of $a_{\uparrow\downarrow}$.
The total numbers of bosons and rungs are denoted by $N$ and $L$, respectively.
We define the particle filling as $f=N/\left(2L\right)$.
Further, we employ the abbreviation $N_\sigma = \sum_{r=0}^{L-1} a_{r,\sigma}^\dagger a_{r,\sigma}$ for the total particle number in leg $\sigma$.
We emphasize that throughout this paper, we consider the value of the flux $\phi/(2\pi) = 1064/769$ that will be most easily realized in the experiment.
In the following, we set $\hbar=1$ and $k_B=1$ and employ the leg-hopping parameter $J$ as our energetic unit of reference.
\subsection{Observables of interest}
\label{subsec:observables}
In the following, we give an account of the observables considered in this paper. They have been chosen due to their experimental relevance in synthetic dimension implementations.
\subsubsection{Momentum distribution functions}
Experimentally, momentum-distribution functions are accessible via time-of-flight measurements.
Moreover, Stern-Gerlach separation allows for leg-resolved measurements~\cite{celi_14, stuhl_15}.
Thus, leg-resolved momentum-distribution functions $n_\sigma(k_m)$ are given by means of momentum operators $\overline{a}_{k_{m},\sigma}^{(\dagger)}$ obtained by Fourier transforming site-local operators $a_{r,\sigma}^{(\dagger)}$ along the spatial dimension,
\begin{align}
n_l\left(k_m\right) &= \left\langle \overline{a}_{k_m,\sigma}^\dagger \overline{a}_{k_m,\sigma} \right\rangle\label{eq:mdf}\\
\overline{a}_{k_m,\sigma} &= \frac{1}{\sqrt{L}}\sum_{r=0}^{L-1} e^{ik_m r}a_{r,\sigma}.
\end{align}
The corresponding quasimomenta read $k_m = 2\pi m/L$ with $m=0,1,\dots,L-1$.
Note that throughout this paper, angled brackets denote ground-state expectation values.
\subsubsection{Chiral current}
Various ground-state phases found in the flux-ladder model have been successfully distinguished by means of their characteristic particle-current patterns (see Ref.~\cite{greschner_16} for an overview).
Particle currents $\left\langle j^\perp_r \right\rangle$ and $\left\langle j^\parallel_{r,\sigma} \right\rangle$, from site $(r,\uparrow)$ to site $(r,\downarrow)$ and from site $(r,\sigma)$ to site $(r+1,\sigma)$, respectively, are obtained from the continuity equation for the occupation of the local lattice sites,
\begin{align}
-\frac{d}{dt}\left\langle n_{r,\sigma} \right\rangle
&= i \left\langle \left[a_{r,\sigma}^\dagger a_{r,\sigma},H\right]\right\rangle \nonumber\\
&= \left\langle j^\parallel_{r,\sigma} \right\rangle - \left\langle j^\parallel_{r-1,\sigma} \right\rangle \pm \left\langle j^\perp_r \right\rangle,
\end{align}
with $\pm$ for $\sigma=\uparrow, \downarrow$.
The corresponding operators read
\begin{align}
j^\parallel_{r,\sigma} &= i J a_{r,\sigma}^\dagger a_{r+1,\sigma} + \mathrm{h.c.},\\
j^\perp_r &= i J_\perp e^{-i r \phi} a_{r,\uparrow}^\dagger a_{r,\downarrow} + \mathrm{h.c.}.
\end{align}
Moreover, the chiral current
\begin{align}
\left\langle j_c \right\rangle = \frac{1}{L-1} \sum_{r = 0}^{L-2} \left( \left\langle j^\parallel_{r,\downarrow} \right\rangle - \left\langle j^\parallel_{r,\uparrow} \right\rangle \right)
\end{align}
represents the particle transport along the legs of the system in opposite directions. 
The parameter dependence of $\left\langle j_c \right\rangle$ can be used to study the vortex-to-Meisner-phase transition in flux ladders~\cite{orignac_01,piraud_15}.
Chiral currents have been experimentally measured by projecting the system into isolated double wells using an optical superlattice and studying its time-dependent dynamics~\cite{atala_14}.
Such schemes become simpler to implement in synthetic flux ladders, where the legs of the ladder correspond to internal spin states that can be imaged independently exploiting Stern-Gerlach separation during time-of-flight expansion~\cite{celi_14, stuhl_15}.
\subsubsection{Leg-population imbalance}
As just mentioned above, in synthetic flux ladders the occupation of the individual legs can be easily determined experimentally because these correspond to different spin states.
Here, we define the leg-population imbalance $\Delta_m$ by means of
\begin{align}
\Delta_m = \frac{ \left| \left\langle N_\uparrow \right\rangle - \left\langle N_\downarrow \right\rangle \right|}{\left\langle N_\uparrow \right\rangle + \left\langle N_\downarrow \right\rangle}.
\end{align}
Note that a finite leg-population imbalance, $\Delta_m>0$, is the key feature of the biased-ladder phase~\cite{wei_14}, which exhibits unequal particle numbers $\left\langle N_\uparrow \right\rangle$ and  $\left\langle N_\downarrow \right\rangle$ in the two legs.
\section{Numerical methods}
\label{sec:numerical_methods}
In this section, we comment on the matrix-product-state based computation of ground states and quench dynamics in the synthetic flux-ladder model. The simulations are performed by means of the SyTen toolkit~\cite{syten, hubigthesis_17}.
The reader primarily interested in the physics may skip this part and jump immediately to Sec.~\ref{sec:gsphases}.
%

%
Throughout our work, the $U(1)$ symmetry associated with the particle-number conservation of the flux-ladder Hamiltonian~\eqref{hamiltonian} is exploited in the matrix-product-state formalism.
In particular, for the calculation of ground states, we employ the single-site variant of the density-matrix renormalization-group method~\cite{white_92,schollwoeck_05,schollwoeck_11}, using subspace expansion~\cite{hubig_15}.
Convergence of the variationally optimized states is ensured by comparing energy expectation values $\left\langle H \right\rangle$, variances $\left\langle H^2 \right\rangle - \left\langle H \right\rangle^2$, as well as all relevant observables, introduced in Sec.~\ref{subsec:observables}, for different values of the site-local bosonic cutoff and for different bond dimensions up to typically $3000$.
We note that for the Hamiltonian model parameters considered in this paper, a truncation to at most six bosons per lattice site is sufficient.
%

%
The quench dynamics presented in Sec.~\ref{sec:quench_dyn} are simulated using the two-site variant of the time-dependent variational principle algorithm~\cite{haegeman_11, paeckel_19}.
Concerning time evolutions, we typically employ bond dimensions of around $500$ and ensure convergence of all relevant observables by increasing the bond dimension and decreasing the time step independently.
\section{Properties of known ground-state phases in the flux-ladder model}
\label{sec:gsphases}
In the presence of interactions, bosonic flux ladders host a panoply of emergent quantum phases~\cite{greschner_16}, which have been studied extensively in the theoretical literature.
In the following, without claiming completeness, we give an account of important ground-state phases of two-leg flux-ladder models.
Readers familiar with the literature on flux ladders might skip this section.
%

%
Orignac and Giamarchi show, in a seminal study based on a bosonization approach, that the minimal two-leg flux ladder exhibits Meissner and vortex-lattice phases, which are reminiscent of a type-II superconductor~\cite{orignac_01}.
The Meissner phase exhibits a homogeneous particle-density profile and uniform particle currents encircling the ladder along its legs in opposite directions.
It is worth noting that the Meissner phase can exist on top of Mott insulators~\cite{petrescu_13, didio_15} as well as on top of superfluids~\cite{piraud_15}.
Typically, the Mott-insulating Meissner phase emerges at a commensurate particle filling per rung. 
It exhibits a central charge $c=0$. 
In the superfluid Meissner phase, the charge gap vanishes and the central charge is given by $c=1$.
%

%
Vortex-lattice phases are regular crystals of localized vortices~\cite{dhar_12, dhar_13, greschner_15}.
In the limit of a vanishing rung hopping and for a homogeneous particle density, a complete devil’s staircase of vortex-lattice phases at each commensurate vortex density is predicted~\cite{orignac_01}.
Interestingly, the breaking of the translational symmetry of the underlying lattice model in the vortex-lattice phases can lead to a reversal of the chiral current~\cite{greschner_15}.
As Meissner phases, vortex-lattice phases can exist on top of superfluids and Mott insulators~\cite{greschner_16}.
However, in general, they are elusive in the strongly interacting regime, requiring weak but finite interaction strengths.
In contrast to vortex-lattice phases, vortex-liquid phases do not exhibit pinned vortices and rung-current correlations.
They show irregular leg-current patterns and can exist for any value of the interaction strength~\cite{piraud_15}.
%

%
Moreover, flux-ladder models host a biased-ladder phase, which was first discussed by Wei and Mueller in 2014~\cite{wei_14}.
The key characteristic of the biased-ladder phase is a finite leg-population imbalance, which spontaneously breaks the leg-inversion symmetry.
It exhibits Meissner-like currents along the legs and vanishing rung currents.
The stability of the biased-ladder phase is typically enhanced by the presence of rung-wise interactions.
%

%
Charge-density waves can be observed in the strongly interacting regime for large values of the magnetic flux~\cite{piraud_15, didio_15}.
Their key feature are particle-density modulations along the legs, while they exhibit homogeneous Meissner-like currents. 
%

%
Finally, we note that precursors of fractional quantum Hall states in bosonic flux ladders have attracted great interest~\cite{grusdt_14, petrescu_15, cornfeld_15, petrescu_17, greschner_17, strinati_17, strinati_19b}.
In general, they require a fine-tuned ratio between the magnetic flux and the particle filling.
\section{Zero-temperature phase diagram}
\label{sec:zeroT_PD}
In the following, we map out the ground-state phase diagram of the synthetic flux-ladder model at particle filling one-half, $f=1/2$, considering SU(2)-symmetric interactions, $U_\uparrow=U_\downarrow=V=U$.
We report on a superfluid as well as a Mott-insulating biased-ladder phase and a Mott-insulating Meissner phase.
\subsection{Overview}
\begin{figure}
	\includegraphics[]{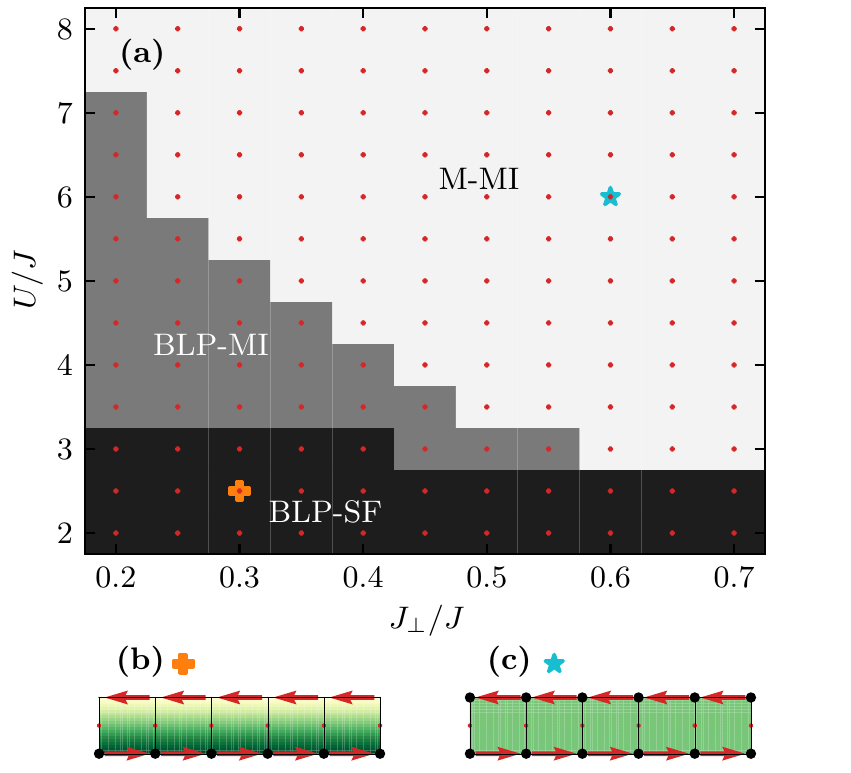}
	\caption{
		Ground-state phase diagram for $f=1/2$ and $U_\uparrow=U_\downarrow=V=U$, featuring superfluid biased-ladder~(BLP-SF), biased-ladder Mott-insulating~(BLP-MI), and Mott-insulating  Meissner~(M-MI) phases.
		(a) Dark gray shading indicates the BLP-SF.
		Light gray shading indicates the BLP-MI.
		Bright regions indicate the M-MI.
		Actual ground states have been computed for the values of $J_\perp$ and $U$ indicated by the red dots, considering ladders with $L=40,~60,~\mathrm{and} ~80$ rungs.
		Note that for noninteracting bosons ($U=0$), the critical value of $J_\perp$ corresponding to the vortex-to-Meissner transition is given by $J_\perp^c/J=4.88$~\cite{tokuno_14, huegel_14}.
		(b) and (c) Local density profile and current patterns in the BLP-SF~($J_\perp/J = 0.3$, $U/J=2.5$) and M-MI~($J_\perp/J = 0.6$, $U/J=6$), respectively.
		The size of the dots and the background shading indicate the local particle density.
		Note the finite leg-population imbalance in (b).
		The red arrows show the local current patterns.
		The data shown in (b) and (c) are for the six most central rungs of a ladder comprising a total number of $L=80$ rungs.
	}
	\label{pd_00}
\end{figure}

Let us start  with Fig.~\ref{pd_00}, which shows the phase diagram as a function of the rung hopping strength $J_\perp$ and the interaction strength $U$.
Within the parameter region spanned by $U/J \in[2,8]$ and $J_\perp/J \in[0.2,0.7]$, extensive density-matrix renormalization-group simulations clearly reveal three kinds of phases, which are also summarized in Table~\ref{tab:gsphases}:
(i) The ground states in the Mott-insulating Meissner phase exhibit uniform particle-density profiles and uniform local current patterns with an effective unit cell comprising one plaquette of the ladder.
Moreover, as shown in Fig.~\ref{chemPot_00}, the central charge of $c=0$ of the Mott-insulating Meissner phase can be well reproduced from the entanglement entropy in the ground state.
(ii) The Mott-insulating biased-ladder phase has a central charge of $c=0$, and, most importantly, it features a finite leg-population imbalance, $\Delta_m>0$.
(iii) The superfluid biased-ladder phase exhibits a finite leg-population imbalance and a central charge of $c=1$, which can also be reproduced from the numerical data.
Note that the local particle currents and particle-density profiles in the superfluid biased-ladder phase and in the Mott-insulating Meissner phase are exemplified in Fig.~\ref{pd_00}(b) and Fig.~\ref{pd_00}(c), respectively.
\subsection{Charge gap and entanglement entropy}
For the purpose of  distinguishing between the Mott-insulating and the superfluid phases, we analyze the charge gap
\begin{align}
	\Delta\mu = \epsilon_{N+1}+\epsilon_{N-1}-2\epsilon_{N}.
\end{align}
Here, $\epsilon_{N}$ denotes the ground-state energy of a setup with $N$ particles and particle filling $f = N/(2L)$.
A vanishing charge gap, $\Delta\mu=0$, indicates a commensurate phase (indicative of a superfluid), while a finite charge gap in the thermodynamic limit, $\lim_{L\rightarrow\infty}\Delta\mu>0$, reveals a Mott insulator.
It should be stressed that $\lim_{L\rightarrow\infty}\Delta\mu$ is estimated by means of a linear extrapolation of finite-size data in $1/L$.
%

%
We emphasize that the presence of rung-wise interactions, $V>0$, generally enhances the stability of the biased-ladder phase.
Interestingly, it has been shown in \cite{zhan_14, barbiero_16} that a finite leg-population imbalance, $\Delta_m>0$, can be found even in the absence of a magnetic field, if the strength of the rung-wise interactions exceeds the site-local interaction strength, $V>U_{\uparrow},U_{\downarrow}$.
However, one does not expect this -- and we have not found any evidence -- for a finite population imbalance at zero flux and close-to SU(2)-symmetric interactions.
%

\begin{figure}
	\includegraphics[]{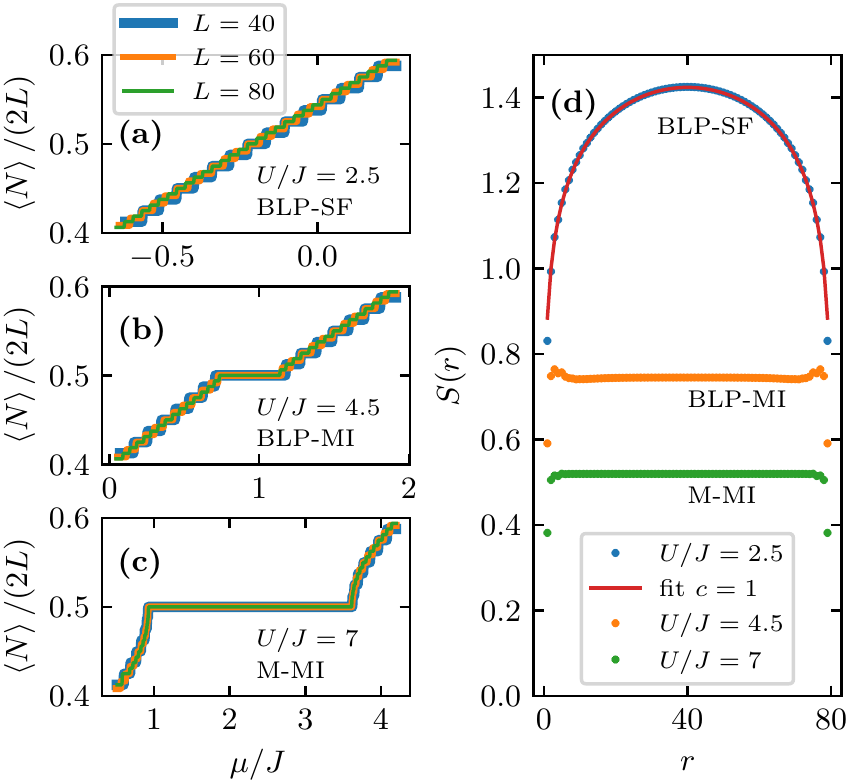}
	\caption{
		Particle number $\left\langle N \right\rangle$ in the grand-canonical ground state versus chemical potential $\mu$ and entanglement spectra, for $J_\perp = 0.3J$ and different values of  $U_\uparrow=U_\downarrow=V=U$.
		$U/J=2.5$ corresponds to the superfluid biased-ladder phase (BLP-SF), $U/J=4.5$ corresponds to the Mott-insulating biased-ladder phase (BLP-MI), and $U/J=7$ corresponds to the Mott-insulating Meissner phase (M-MI).
		(a) $\left\langle N \right\rangle$ versus $\mu$, $U=2.5J$, BLP-SF.
		Data are shown for $L=40,60,$ and $80$ rungs.
		(b) $U=4.5J$, BLP-MI.
		(c) $U=7J$, M-MI.
		The plateaus in (b) and (c) indicate the appearance of Mott-insulators at filling $f=1/2$; see also Fig.~\ref{pd_U_00}.
		(d) Entanglement entropy $S(r)$ obtained for bipartitions corresponding to cuts between rung $(r - 1)$ and rung $r$ for $U = 2.5J$~(BLP-SF, $c=1$), $U = 4.5J$~(BLP-MI, $c=0$), and $U=7J$~(M-MI, $c=1$); considering $f = 1/2$.
		The red dashed line has been obtained by fitting the parameter $g$ in Eq.~\eqref{entanglement_scaling} to the $U = 2.5J$ data, considering $c=1$.
	}
	\label{chemPot_00}
\end{figure}

%
In Fig.~\ref{chemPot_00}, we present additional results underlining the ground-state phase diagram for SU(2)-symmetric interactions $U_\uparrow=U_\downarrow=V=U$ and particle filling $f=1/2$ presented in Fig.~\ref{pd_00}.
Figure~\ref{chemPot_00}(a), Fig.~\ref{chemPot_00}(b), and Fig.~\ref{chemPot_00}(c) show the particle number $\left\langle N \right\rangle$ in the grand-canonical ground state as a function of the chemical potential $\mu$ for $U/J=2.5$, $U/J=4.5$, and $U/J=7$, respectively, considering $J_\perp/J=0.3$.
Note that these parameters are also considered in Fig.~\ref{pd_U_00}.
As discussed above, for $U/J=4.5$ and $U/J=7$, Mott-insulating ground states appear at filling $f=1/2$.
The plateaus at $\left\langle N \right\rangle = L$ in the $\left\langle N \right\rangle$ versus $\mu$ curves shown in Fig.~\ref{chemPot_00}(b) and Fig.~\ref{chemPot_00}(c) are indicative for these Mott-insulating phases.
Figure~\ref{chemPot_00}(d) shows the entanglement entropy $S(r)$ as obtained for bipartitions corresponding to cuts between rung $(r - 1)$ and rung $r$, $U=2.5J$ (biased-ladder superfluid, $c=0$), $U=4.5J$ (biased-ladder Mott insulator, $c=0$), and $U=7J$ (Mott-insulating Meissner phase, $c=1$), considering a particle filling $f=1/2$.
Note that the ground-state entanglement entropy is predicted to scale as~\cite{vidal_03,calabrese_04}
\begin{align}
S(r) = \frac{c}{6}\log\left(\frac{L}{\pi}\sin\left(\frac{\pi r}{L}\right)\right)+g\label{entanglement_scaling},
\end{align}
with $c$ being the central charge and $g$ a nonuniversal constant.
The red line is obtained by least-square fitting the offset $g$ in the expression above to the $U/J=2.5$ data, considering a central charge $c=1$, which is expected for the superfluid biased-ladder phase.
\subsection{Momentum-distribution functions}
\begin{figure}
	\includegraphics[]{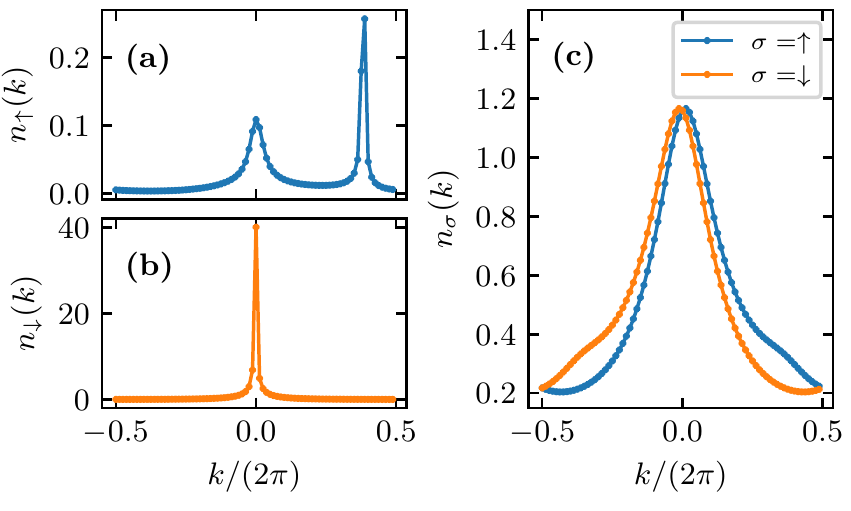}
	\caption{
		Leg-resolved momentum-distribution function $n_\sigma(k)$, $f=1/2$, $L=80$.
		(a) and (b) are for the $\uparrow$ and $\downarrow$ leg, respectively, considering $J_\perp/J=0.3$ and $U_\uparrow=U_\downarrow=V=2.5J$,  corresponding to the biased-ladder superfluid phase.
		(c) $J_\perp/J=0.6$ and $U_\uparrow=U_\downarrow=V=6J$, Mott-insulating Meissner phase.
		Note that in the Meissner phase, one finds  $n_\uparrow(k)=n_\downarrow(-k)$, which does not apply to the biased-ladder phase.
		The current patterns and density profiles for the parameters considered in this figure are shown in Fig.~\ref{pd_00}(b) and Fig.~\ref{pd_00}(c).
	}
	\label{moDist_00}
\end{figure}

In Fig.~\ref{moDist_00}, we exemplify leg-resolved  ($\sigma=\uparrow, \downarrow$) momentum-distribution functions $n_\sigma(k)$, as defined in Eq.~\eqref{eq:mdf}, in the biased-ladder superfluid phase, and in the Mott-insulating Meissner phase.
Figure~\ref{moDist_00}(a) and Fig.~\ref{moDist_00}(b) show $n_\uparrow(k)$ and $n_\downarrow(k)$  in the superfluid biased-ladder phase, for $J_\perp/J=0.3$ and $U/J=2.5$.
Note that for the considered parameters, $\left\langle N_\downarrow \right\rangle \gg \left\langle N_\uparrow \right\rangle$ and the maximum values of $n_\uparrow(k)$ and $n_\downarrow (k)$ differ by two orders of magnitude.
Also, $n_\downarrow(k)$ is sharply peaked around zero quasi-momentum, while the displaced peak in $n_\uparrow(k)$ is a signature of the superfluid biased-ladder phase.
Figure~\ref{moDist_00}(c) focuses on the Mott-insulating Meissner phase and shows $n_\sigma(k)$ as obtained for $J_\perp/J=0.6$ and $U=6J$.
Note that in the Meissner phase, the leg-resolved momentum-distribution functions fulfill the symmetry relation $n_\uparrow(k)=n_\downarrow(-k)$.
Moreover, both momentum-distribution functions, $n_\uparrow(k)$ and $n_\downarrow(k)$, exhibit peaks in the immediate proximity to $k=0$.
Note that the current patterns and density profiles for the parameters considered in Fig.~\ref{moDist_00} are presented in Fig.~\ref{pd_00}(b) and Fig.~\ref{pd_00}(c).
\subsection{Tuning the rung hopping strength}
\begin{figure}
	\includegraphics[]{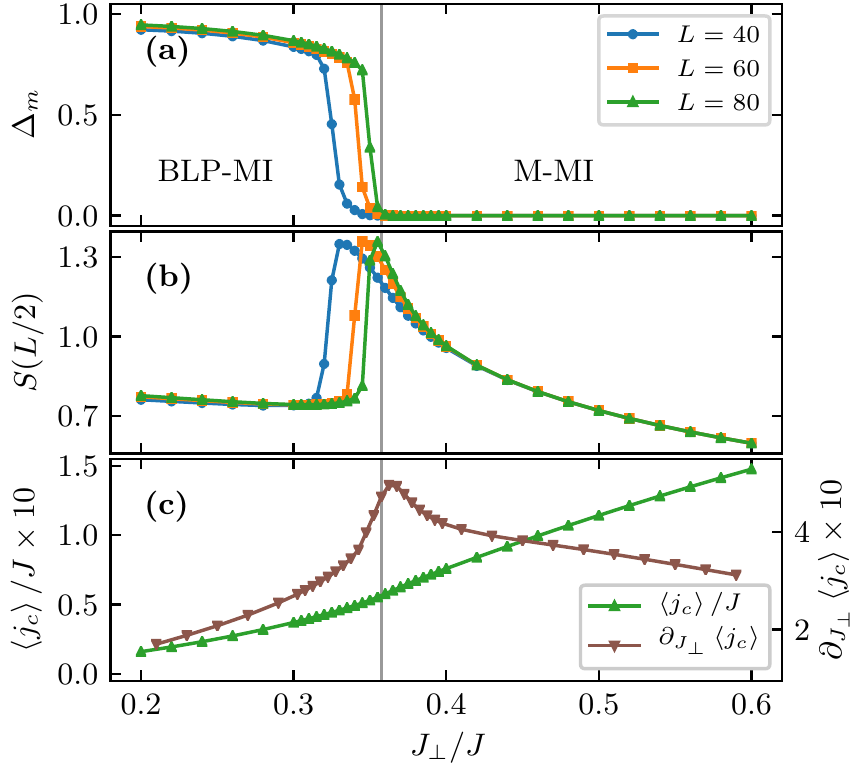}
	\caption{
		Biased-ladder Mott insulator~(BLP-MI) and Mott-insulating Meissner phase~(M-MI) for $U_\uparrow=U_\downarrow=V=4.5J$, and $f=1/2$.
		(a) Leg-population imbalance $\Delta_m$ versus $J_\perp$ considering $L=40,60,$ and $80$ rungs.
		(b) Entanglement entropy $S(L/2)$ for a bipartition corresponding to a cut between the two most central rungs of the ladder.
		Note that the legend from panel (a) also applies to panel (b).
		(c) Chiral current $\left\langle j_c \right\rangle$ and slope of the chiral current $\partial_{J_\perp} \left\langle j_c \right\rangle$ versus $J_\perp$ for $L=80$ rungs.
	}
	\label{pd_JP_00}
\end{figure}

Next, we concentrate on a horizontal cut through the phase diagram introduced in Fig.~\ref{pd_00} at $U/J=4.5$ and elucidate the biased-ladder-Mott-insulator-to-Mott-insulating Meissner phase transition in Fig.~\ref{pd_JP_00}.
Figure~\ref{pd_JP_00}(a) shows the ground-state leg-population imbalance $\Delta_m$ as a function of $J_\perp$ for systems with $L=40,60,\text{ and }80$ rungs.
The difference between the $L=60$ and $L=80$ data is almost negligible on the scale of the figure and the abrupt change of $\Delta_m$ clearly reveals the locus of the phase transition, which is also indicated by the vertical gray line.
The half-cut entanglement entropy $S(L/2)$, which corresponds to a bipartition between the two most central rungs of the ladder, indicates the biased-ladder to Meissner-phase transition; see Fig.~\ref{pd_JP_00}(b).
Within the considered region $J_\perp/J\in\left[0.2,0.6\right]$, the  chiral current $\left\langle j_c\right\rangle$ increases monotonically with the rung hopping strength $J_\perp$, which can also be seen in Fig.~\ref{pd_JP_00}(c).
However, a kink in $\left\langle j_c \right\rangle$ marks the point of the  biased-ladder to Meissner-phase transition. This kink is also evident in the derivative $\partial_{J_\perp}\left\langle j_c \right\rangle$, as  illustrated in Fig.~\ref{pd_JP_00}(c).
\subsection{Tuning the interparticle interaction strength}
\begin{figure}
	\includegraphics[]{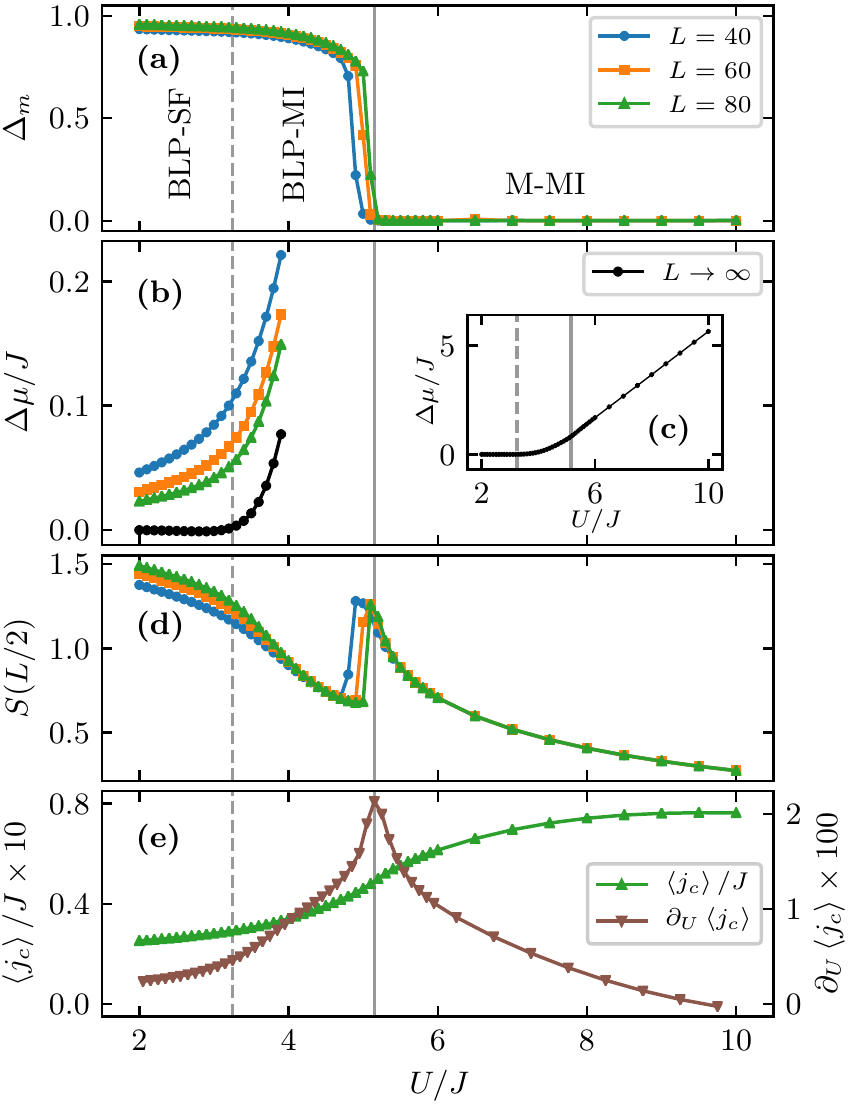}
	\caption{
		Biased-ladder superfluid phase~(BLP-SF), biased-ladder Mott insulator~(BLP-MI), and Mott-insulating Meissner phase~(M-MI) for $U=U_\uparrow=U_\downarrow=V$, $J_\perp/J=0.3$, and $f=1/2$.
		(a) Leg-population imbalance $\Delta_m$ versus $U$ considering $L=40,60,80$ rungs.
		The vertical gray lines indicate the estimated locus of the quantum phase transitions.
		(b) Charge gap $\Delta\mu$ versus $U$.
		The black solid line shows the extrapolated value of $\Delta\mu$ in the thermodynamic limit, $\lim\limits_{L\rightarrow\infty}\Delta\mu$.
		The inset (c) shows the charge gap in the thermodynamic limit for $U/J \in \left[ 2, 10 \right]$.
		(d) Entanglement entropy $S(L/2)$ for a bipartition corresponding to a cut between the two most central rungs of the ladder.
		Note that the legend from panel (a) also applies to panel (b) and (d).
		(e) Chiral current $\left\langle j_c \right\rangle$ and slope of the chiral current $\partial_{U} \left\langle j_c \right\rangle$ versus $U$ considering $L=80$ rungs.
		}
\label{pd_U_00}
\end{figure}

Figure~\ref{pd_U_00} focuses on a vertical cut through the phase diagram presented in Fig.~\ref{pd_00} at $J_\perp/J=0.3$.
The abrupt change of the population imbalance when increasing $U$ above approximately $5J$, shown in Fig.~\ref{pd_U_00}(a), pinpoints the transition from the Mott-insulating biased-ladder phase to the Mott-insulating Meissner phase.
Note that finite-size effects in the population imbalance for systems with more than $L=60$ rungs are negligible on the scale of the figure.
Most interestingly, the system undergoes a superfluid to Mott-insulator transition within the biased-ladder phase when increasing $U$ above approximately $3.2J$.
This is revealed by the opening of a charge gap $\Delta\mu$, as shown in Fig.~\ref{pd_U_00}(b).
In particular, in Fig.~\ref{pd_U_00}(b), we plot $\Delta\mu$ for systems with $L=40,60,\text{ and } 80$ rungs (colored lines) as well as $\lim_{L\rightarrow\infty}\Delta\mu$ (black line), which has been obtained using a linear extrapolation of the finite-size data in $(1/L)$, as discussed above.
Figure~\ref{pd_U_00}(c) shows the extrapolated charge gap $\lim_{L\rightarrow\infty}\Delta\mu$ for $U/J\in[2,10]$.
The half-cut entanglement entropy  $S(L/2)$ shown in Fig.~\ref{pd_U_00}(d) exhibits a discontinuity at the transition from the Mott-insulating biased-ladder phase to the Mott-insulating Meissner phase.
Moreover, $S(L/2)$ is independent of the system size $L$ in the Mott-insulating phases, while it shows a dependence on $L$ in the superfluid biased-ladder phase~\cite{eisert_10, greschner_16}.
The chiral current $\left\langle j_c\right\rangle$ is shown in Fig.~\ref{pd_U_00}(e) as a function of $U$.
In analogy to results presented in Fig.~\ref{pd_JP_00}(c), a maximum in its slope $\partial_U \left\langle j_c\right\rangle$ indicates the biased-ladder to Meissner-phase transition.
Note that the vertical gray lines show the estimated points of the quantum-phase transitions.
\subsection{Effect of an additional trapping potential}
\begin{figure}
	\includegraphics[]{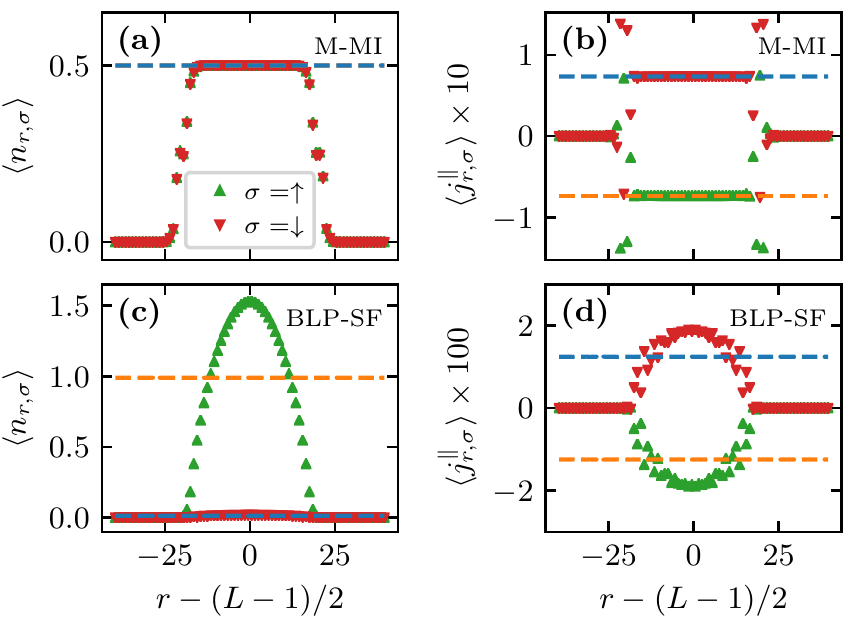}
	\caption{
		Effect of a quadratic trapping potential.
		The figure shows ground-state particle-density profiles $\left\langle n_{r,\sigma}\right\rangle$ [(a), (c)] and local leg currents $\langle j_{r,\sigma}^\parallel \rangle$ [(b), (d)] in the central one-half of the ladder.
		We consider a harmonic trapping potential as given in Eq.~\eqref{eq:trap} with $\mu_t/J=60$, $L=160$, $N=40$ bosons, and  $U=U_\uparrow=U_\downarrow=V$.
		(a) and (b) are for $U/J=6$ and $J_\perp/J=0.6$, which corresponds to the Mott-insulating Meissner phase (M-MI) in the absence of the trapping potential and for a particle filling $f=1/2$; see Fig.~\ref{pd_00}.
		(c) and (d) are for $U/J=2.5$ and $J_\perp/J=0.3$, corresponding to the superfluid biased-ladder phase (BLP-SF).
		In all panels, the green upper triangles and the red lower triangles are for the $\sigma=\uparrow$ leg and the $\sigma=\downarrow$ leg, respectively.
		The dashed orange ($\sigma=\uparrow$) and blue ($\sigma=\uparrow$) lines show results in the absence of the harmonic potential and for a particle filling $f=1/2$.
		Note that in panel (a), the data for $\sigma=\uparrow$ and $\sigma=\downarrow$ are on top of each other.
	}
	\label{trap}
\end{figure}
In the experimental implementation of the flux-ladder model proposed in Sec.~\ref{sec:model:exp}, the atoms are captured by a harmonic trapping potential.
Hence, in Fig.~\ref{trap} we show particle-density and leg-current profiles for ground states of the flux-ladder Hamiltonian~\eqref{hamiltonian} in the presence of an additional quadratic potential given by
\begin{align}
	V_t = {\mu_t} \sum_{\sigma=\uparrow,\downarrow} \sum_{r=0}^{L-1} \frac{ \left(r-(L-1)/2\right)^2}{\left((L-1)/2\right)^2} n_{r,\sigma}\,.\label{eq:trap}
\end{align}
Concretely, we consider a ladder with ${L=160}$ rungs, ${N=40}$ bosons, and ${\mu_t/J=60}$.
Due to the effect of the quadratic potential, the particles localize in the center of the system. 
For ${U/J=6}$ and ${J_\perp/J=0.6}$, one finds a Mott region in the central one-quarter of the ladder with a homogeneous particle density ${\left\langle n_{r,\sigma} \right\rangle=0.5}$ and homogeneous leg currents $\langle j^{\parallel}_{r,\sigma} \rangle$, as can be seen by the triangle symbols in Fig.~\ref{trap}(a) and Fig.~\ref{trap}(b).
It is worth noting that in the absence of the trapping potential and for a particle filling $f=1/2$, the considered values of $J_\perp$ and $U$ correspond to the Mott-insulating Meissner phase.
Also, Fig.~\ref{trap}(b) shows that the local leg currents in the Mott region are in accordance with the leg currents observed in the absence of a trapping potential for $f=1/2$, which are indicated by the dashed lines. 
For ${U/J=2.5}$ and ${J_\perp/J=0.3}$, one observes a finite population imbalance in the center of the system, where the particles accumulate, see Fig.~\ref{trap}(c).
This is in accordance with the superfluid biased-ladder phase, which is found for the considered values of $J_\perp$ and $U$ in the absence of a trapping potential and for a particle filling $f=1/2$.
Figure~\ref{trap}(d) shows that for ${U/J=2.5}$ and ${J_\perp/J=0.3}$, one finds symmetric Meissner-like leg currents in the center of the system.
The leg currents observed in the superfluid biased-ladder ground state at $f=1/2$ and in the absence of a trapping potential are indicated by the dashed lines and shown for comparison.
We conclude that the relevant ground-state phases discussed in this paper can be observed in the presence of a strong trapping potential.
\section{Quench dynamics}
\label{sec:quench_dyn}
Preparing a system close to its ground state experimentally can be notoriously difficult.
Therefore, it is highly desirable to develop practical schemes to explore the various phases existing in the interacting flux-ladder model~\cite{greschner_16}.
This is underlined by recent experimental advances in noninteracting ladders, where elaborate loading procedures enabled the observation of chiral edge states~\cite{mancini_15,stuhl_15} and the estimation of Chern numbers~\cite{mugel_17, genkina_19}.
\begin{figure}
	\includegraphics[]{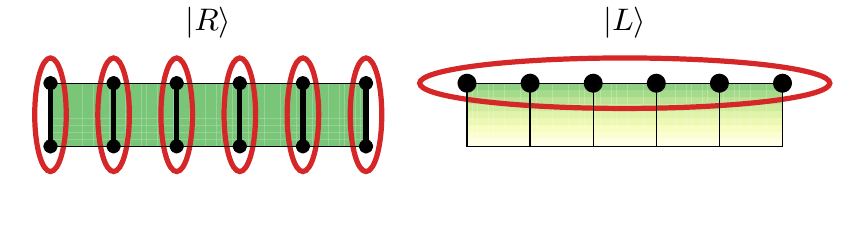}
	\caption{
		Sketch of the initial states $\ket{R}$ and $\ket{L}$ considered in quantum quenches.
		$\ket{R}$ represents the one-particle-per-rung ground state of the ladder Hamiltonian $H$, introduced in Eq.~\eqref{hamiltonian}, for vanishing leg hopping, $J=0$.
		Note that each rung is occupied by exactly one particle.
		$\ket{L}$ represents the $L$-particles-on-one-leg ground state of $H$ for vanishing rung hopping, $J_\perp=0$.
	}
	\label{init_states}
\end{figure}
%

%
Here, we present feasible quench protocols which might allow one to probe the chiral current in the interacting Meissner phase and to detect signatures of an underlying biased-ladder phase in the transient dynamics of the leg-population imbalance.
In Sec.~\ref{sec:quench_dyn:R}, our focus is on the chiral current in the Meissner phase.
There, we study the instantaneous turning on of leg hopping in the synthetic flux-ladder model, considering a rung-localized initial state, which is here denoted by $\ket{R}$.
Explicitly, for a vanishing leg hopping ($J=0$), $\ket{R}$ is the one-particle-per-rung ground state of the Hamiltonian $H$ introduced in Eq.~\eqref{hamiltonian}.
It is sketched in Fig.~\ref{init_states} and given by
\begin{align}
\ket{R} = 2^{-L/2} \prod_{r=0}^{L-1} \left(e^{-ir\phi/2}a_{r,\uparrow}^{\dagger}+e^{ir\phi/2}a_{r,\downarrow}^{\dagger}\right)\ket{\mathrm{vac}},
\end{align}
where $\ket{\mathrm{vac}}$ denotes the vacuum state with $\left\langle N \right\rangle=0$.
In Sec.~\ref{sec:quench_dyn:L}, we concentrate on the leg-population imbalance and investigate the instantaneous turning on of rung hopping considering a leg-localized initial state $\ket{L}$.
Here, $\ket{L}$ represents the  $L$-particles-on-one-leg ground state of $H$, as obtained for vanishing rung hopping ($J_\perp=0$).
We recall that both initial states are experimentally accessible and the considered quench schemes are realistic in current quantum-gas platforms, see Sec.~\ref{sec:model:exp}.
\subsection{Probing the chiral current}
\label{sec:quench_dyn:R}
\begin{figure}
	\includegraphics[]{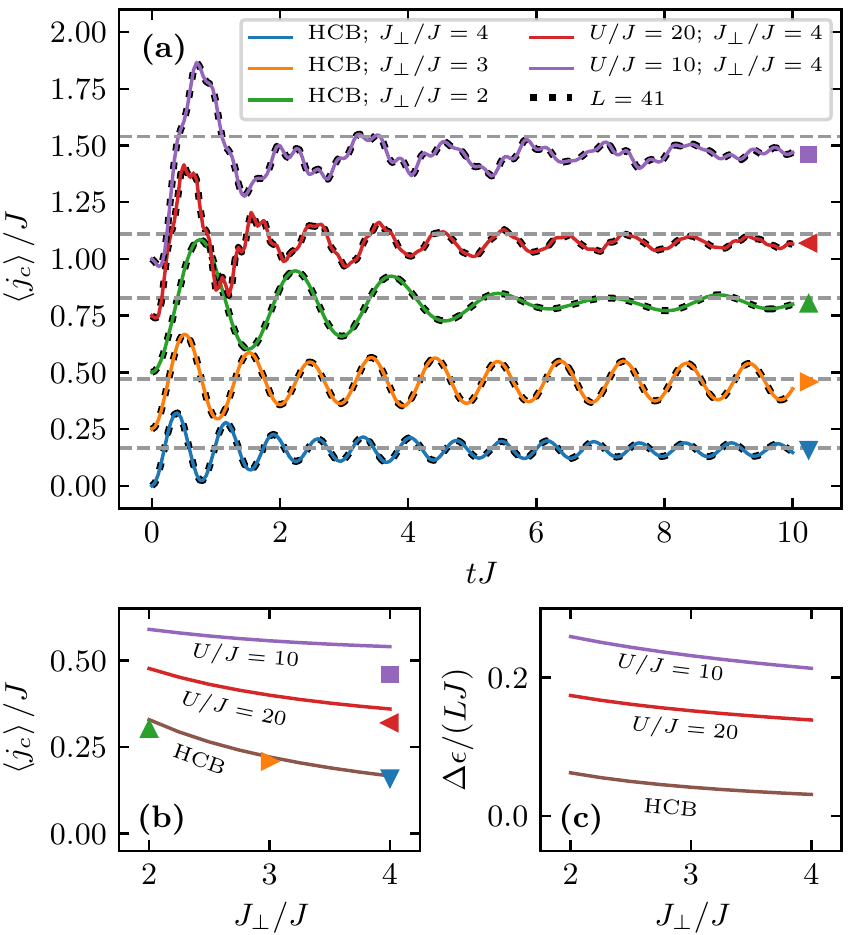}
	\caption{
		Transient dynamics in the chiral current $\left\langle j_c \right\rangle$ after the instantaneous turning on of leg hopping for the  initial state $\ket{R}$ and $V=0$, $U_\uparrow=U_\downarrow=U$, $L=61$.
		(a) $\left\langle j_c \right\rangle$ versus time $t$.
		HCB refers to hard-core bosons and the  data corresponding to different parameters ($J_\perp$ and $U$) are vertically offset by $0.25$, $0.5$, $0.75$, and $1$, for the purpose of a clear presentation.
		The black dotted lines are for $L=41$ rungs, showing the negligible role of finite-size effects.
		In order to neglect boundary effects, $\left\langle j_c \right\rangle$ is computed in the central one third of the ladder.
		Colored solid symbols show the time-averaged value of $\left\langle j_c\right\rangle$ for $tJ\in\left[4,10\right]$.
		Grey dashed lines show $\left\langle j_c\right\rangle$ in the ground state of the post-quench Hamiltonian.
		(b) Overview over the $J_\perp$-dependence of $\left\langle j_c \right\rangle$ in the ground state of the post-quench Hamiltonian~(colored solid lines).
		Colored symbols depict the time-averaged values of $\left\langle j_c \right\rangle$, which are also shown in (a).
		(c) Quench energy $\Delta\epsilon$ versus $J_\perp$.
	}
	\label{dyn_chiral_curr_00}
\end{figure}

Let us start  with the presentation of quench results in the presence of site-local interactions but without rung-wise interactions, $U_\uparrow=U_\downarrow=U$ and $V=0$.
Figure~\ref{dyn_chiral_curr_00}(a) shows the transient dynamics in the chiral current $\left\langle j_c \right\rangle$, which are  induced by the instantaneous turning on of leg hopping in the rung-localized initial state $\ket{R}$.
Explicitly, time evolutions of $\left\langle j_c \right\rangle$ are shown for hard-core bosons (HCB), considering $J_\perp/J = 4,3$, and $2$,  as well as for finite interaction strengths, $U/J=20$ and $U/J=10$, considering $J_\perp/J = 4$.
For the purpose of a clear presentation, the data corresponding to different values of $U$ and $J_\perp$ are vertically offset by $m \times 0.25$, with $m=0,1,\dots,4$.
In order to neglect boundary effects, $\left\langle j_c \right\rangle$ is computed in the central one third of the ladder.
A comparison of the $L=61$ data (solid colored lines) and the $L=41$ data (black dotted lines) reveals that finite-size effects are negligible within the considered time interval $tJ\in[0,10]$.
Most interestingly, Fig.~\ref{dyn_chiral_curr_00}(a) shows that after the instantaneous turning on of leg hopping, $\left\langle j_c \right\rangle$ oscillates around a finite value.
The colored solid symbols represent the time averages of $\left\langle j_c \right\rangle$, which are computed in the interval $tJ\in[4,10]$.
Remarkably, for strong interactions, theses time averages exhibit a very similar dependence on the model parameters as $\left\langle j_c \right\rangle$ in the ground state of the post-quench Hamiltonian, which are indicated by the horizontal dashed gray lines.

Figure~\ref{dyn_chiral_curr_00}(b) gives an overview of the $J_\perp$-dependence of $\left\langle j_c \right\rangle$ in the ground state for $U/J=10, 20$ and hard-core bosons, including  the time averages from Fig.~\ref{dyn_chiral_curr_00}(a).
We emphasize that for $U/J=10$ and $U/J=20$ as well as for hard-core bosons, the vortex-to-Meissner transition appears for values of $J_\perp/J<2$~\cite{piraud_15}.
Thus, all of the parameters considered in Fig.~\ref{dyn_chiral_curr_00} correspond to the Meissner phase.
The quench energy $\Delta\epsilon$ measures the difference between the energy in the flux-ladder after turning on the hopping elements and  the ground-state energy of the post-quench Hamiltonian $H$ introduced in Eq.~\eqref{hamiltonian}.
Explicitly, it is given by
\begin{align}
	\Delta\epsilon = \bra{\psi} H \ket{\psi} - \epsilon,
\end{align}
where $\epsilon$ is the ground-state energy and $\ket{\psi} = \ket{R}, \ket{L}$ is the considered initial state.
Figure~\ref{dyn_chiral_curr_00}(c) elucidates that $\Delta\epsilon$ decreases with increasing interaction strength and increasing rung hopping strength $J_\perp$.
This is in accordance with our observation that the quench protocol discussed here reproduces the chiral current in the ground state especially well in the strongly interacting and large-$J_\perp$ regime, which is deep in the Meissner phase.
Lastly, we note that even though the case of vanishing rung-wise interactions ($V=0$) is not particularly relevant in synthetic dimension implementations, it is still of general interest as it represents a variant of the flux-ladder model which has been extensively studied in previous works, see, for instance, Refs.~\cite{orignac_01,piraud_15,greschner_16}.
%

\begin{figure}
	\includegraphics[]{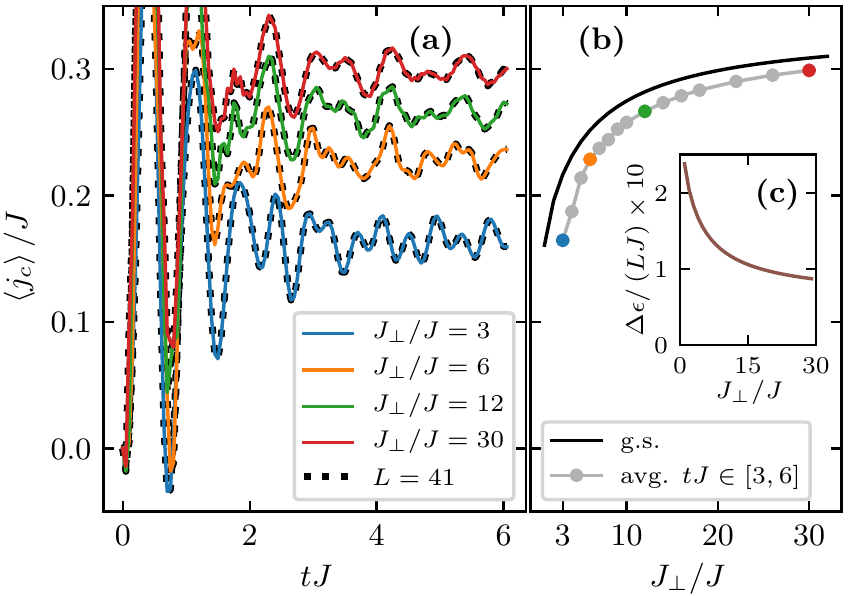}
	\caption{
		Transient dynamics in the chiral current $\left\langle j_c \right\rangle$ after the instantaneous turning on  of leg hopping for various values of $J_\perp$, the initial state $\ket{R}$, $U_\uparrow=U_\downarrow=V=8J$, and $L=61$.
		(a) $\left\langle j_c \right\rangle$ versus time  $t$ for $J_\perp/J=3,6,12$ and $30$.
		The black dotted lines are for $L=41$ rungs, showing the negligible role of finite-size effects.
		(b) The black solid line depicts $\left\langle j_c\right\rangle$ versus $J_\perp$ in the ground state of the post-quench Hamiltonian.
		Gray symbols indicate the time-averaged values of $\left\langle j_c \right\rangle$ considering $tJ\in[5,10]$ after the quench.
		Note that the values of $\left\langle j_c \right\rangle$ for which the transient dynamics are shown in panel (a) are highlighted by the corresponding colors.
		The inset (c) shows the quench energy $\Delta\epsilon$ versus $J_\perp$.
		In order to neglect boundary effects, $\left\langle j_c \right\rangle$ is computed in the central third of the ladder.
	}
	\label{dyn_chiral_curr_01}
\end{figure}

%
Next, we concentrate on SU(2)-symmetric interactions, which are especially relevant in ladders realized by means of a synthetic dimension, such as the $^{41}$K system proposed in Sec.~\ref{sec:model:exp}.
In Fig.~\ref{dyn_chiral_curr_01}, we consider the instantaneous turning on of leg hopping for $U_\uparrow = U_\downarrow = V = 8J$ in a system with $L=61$ rungs and for model parameters corresponding to the Meissner phase.
Figure~\ref{dyn_chiral_curr_01}(a) shows the transient dynamics in $\left\langle j_c \right\rangle$, which are computed in the central third of the ladder, for different values of the rung hopping strength $J_\perp/J = 3,6,12$, and $30$ (solid colored lines).
In analogy to Fig.~\ref{dyn_chiral_curr_00}, it can be seen that after an initial transient regime up to time $tJ\approx3$, $\left\langle j_c \right\rangle$ oscillates around a finite value.
Moreover, the black dashed lines, which are for $L=41$ and on top of the $L=61$ results, suggest a negligible influence of boundary effects for $ tJ \in[0,6]$.
%

%
We emphasize that the time-averaged values of $\left\langle j_c \right\rangle$ in the interval $tJ\in[3,6]$ provide a good estimate for $\left\langle j_c \right\rangle$ in the ground state of the post-quench Hamiltonian, capturing the $J_\perp$-dependence of the latter.
This is elucidated  in Fig.~\ref{dyn_chiral_curr_01}(b) for various values of $J_\perp/J\in[3,30]$, where the solid black line corresponds to the ground state and the gray symbols depict the time averages.
Note that time-averaged chiral currents for which the transient dynamics are shown in Fig.~\ref{dyn_chiral_curr_01}(a) are highlighted by the corresponding colors in Fig.~\ref{dyn_chiral_curr_01}(b).
Additionally, Fig.~\ref{dyn_chiral_curr_01}(c) reveals that the quench energy $\Delta\epsilon$ decreases with increasing $J_\perp$, suggesting that the quench protocol is especially useful in the regime of strongly coupled legs, which is deep in the Meissner phase.
Finally, we conclude that after the instantaneous turning on of leg hopping in the Meissner phase, the chiral current in the short-time dynamics exhibits a similar dependence on $J_\perp$ as the chiral current in the corresponding ground state.
\subsection{Signatures of the biased-ladder phase}
\label{sec:quench_dyn:L}
\begin{figure}
	\includegraphics[]{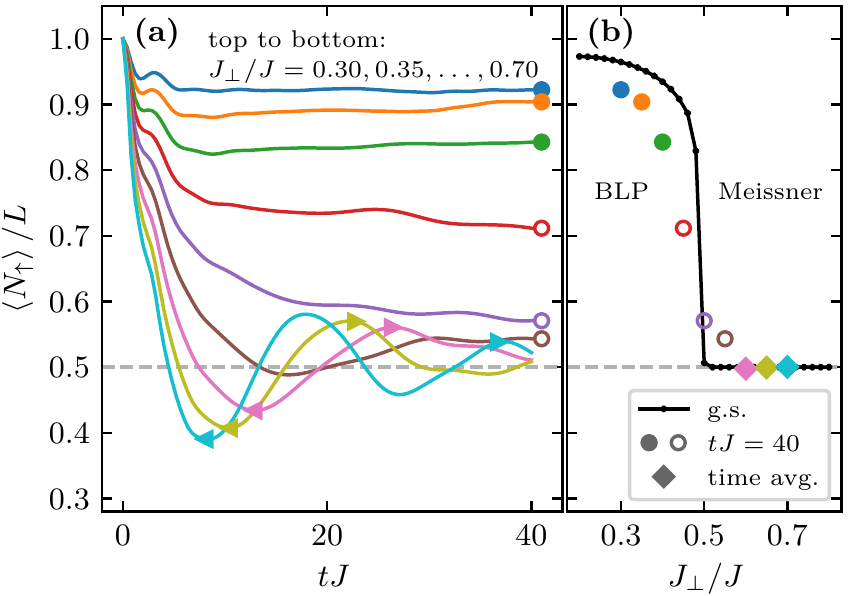}
	\caption{
		Transient dynamics in the leg-population imbalance after the instantaneous turning on of rung hopping for the initial state $\ket{L}$, $U_\uparrow=U_\downarrow=V=3.5J$, $L=61$, and $J_\perp/J \in [0.3,0.7]$.
		(a) Particle number $\left\langle N_\uparrow \right\rangle / L$ in the $\uparrow$-leg versus time $t$.
		The gray dashed line indicates a vanishing leg-population imbalance, corresponding to $\left\langle N_{\uparrow} \right\rangle /L = 0.5$ (note that there is a total number of $\left\langle N \right\rangle=L$ particles).
		For values of $J_\perp$ that show oscillations of $\left\langle N_\uparrow \right\rangle$ in $tJ\in[0,40]$, left and right triangles indicate the time intervals for which the time-averaged data (diamonds) shown in (b) are computed.
		Concerning parameters, for which $\left\langle j_c \right\rangle$ does not show clear oscillations in the considered time interval, we consider the values attained at $tJ=40$ (solid and open circles).
		(b) $\left\langle N_\uparrow \right\rangle/L $ versus $J_\perp$  in the ground state of the post-quench Hamiltonian (black line), indicating the biased-ladder phase (BLP) and the Meissner phase.
		Symbols depict the time-averaged values of $\left\langle N_\uparrow \right\rangle$ (diamonds) or  $\left\langle N_\uparrow \right\rangle$ at $tJ=40$ (solid and open circles).
		}
	\label{dyn_pop_imb_00}
\end{figure}

In Fig.~\ref{dyn_pop_imb_00}, we focus on the instantaneous turning on of rung hopping in the leg-localized initial state $\ket{L}$.
It is shown that signatures of an underlying biased-ladder phase of the post-quench Hamiltonian can be observed in the transient dynamics of the density imbalance between the legs of the ladder.
Here, the considered model parameters are $U_\uparrow=U_\downarrow=V=3.5J$ and $J_\perp/J\in [0.3,0.7]$.
We stress that for $J_\perp/J<0.5$, the ground state of the post-quench Hamiltonian is in the biased-ladder phase, while $J_\perp/J>0.5$ corresponds to the Meissner phase, see Fig.~\ref{pd_00}.
Figure~\ref{dyn_pop_imb_00}(a) depicts the average particle number $\left\langle N_\uparrow \right\rangle / L$ in the $\uparrow$-leg versus the time $t$ elapsed after the quench.
Most interestingly, for parameters which clearly correspond to the biased-ladder phase, namely $J_\perp/J=0.3$,  $J_\perp/J=0.35$, and $J_\perp/J=0.4$, a very stable density imbalance is maintained throughout the considered time interval $tJ\in[0,40]$.
On the other hand, for the parameters corresponding to the Meissner phase, $J_\perp/J=0.60$,  $J_\perp/J=0.65$, and $J_\perp/J=0.70$, $\left\langle N_\uparrow \right\rangle$ quickly starts to decay and to oscillate around $L/2$, which corresponds to a vanishing leg-population imbalance characteristic of the Meissner phase.
For values of $J_\perp$ which are in the immediate proximity to the biased-ladder to Meissner-phase transition, namely $J_\perp/J=0.45$,  $J_\perp/J=0.5$, and $J_\perp/J=0.55$, $\left\langle N_\uparrow \right\rangle$ does not exhibit decaying oscillations in the considered time interval $tJ\in[0,40]$ but suggests a possible decay towards $L/2$ on an intermediate time scale.
Figure~\ref{dyn_pop_imb_00}(b) shows  $\left\langle N_\uparrow \right\rangle$ in the ground state of the post-quench Hamiltonian (black line), which unambiguously indicates the biased-ladder phase for $J_\perp/J<0.5$.
The diamond-shaped symbols depict time-averages of $\left\langle N_\uparrow \right\rangle$, considering intervals as indicated by the left and right triangles in Fig.~\ref{dyn_pop_imb_00}(a).
They reveal that for model parameters corresponding to the Meissner phase, the density imbalance between the legs quickly vanishes after the turning on of rung hopping.
However, for values of $J_\perp$ corresponding to the biased-ladder phase, the values of $\left\langle N_\uparrow \right\rangle$ attained at $tJ=40$, which are indicated by the solid circles, reveal a residual finite density imbalance between the legs on the time scales simulated here.
Since the quench puts the system at finite temperature, we expect that the imbalance will ultimately decay to zero, consistent with studies of order-parameter decays in other one-dimensional systems~\cite{Barmettler2009,Enss2012,Bauer2015}.
Moreover, there is a $J_\perp$-dependence that is in accordance with the one of the ground state in the post-quench Hamiltonian.
Similarly, for values of $J_\perp$ in the immediate proximity to the biased-ladder to Meissner-phase transition, the values of $\left\langle N_\uparrow \right\rangle$ attained at $tJ=40$ are indicated by the open circles.
We conclude that the underlying biased-ladder phase leaves signatures in the short-time dynamics following the instantaneous turning on of rung hopping in the leg-localized initial state $\ket{L}$.
\section{Summary}
\label{sec:summary}
In this paper, we studied the ground-state phases and quench dynamics in an interacting bosonic flux-ladder model.
The focus was on model parameters and specifics that are realistic in a $^{41}$K setup which exploits two internal atomic states ($\uparrow$, $\downarrow$)  as a synthetic dimension.
Explicitly, we concentrated on rung-wise SU(2)-symmetric interactions $U_\uparrow=U_\downarrow=V=U$, a particle filling of one boson per rung $f=1/2$,  and a value of the magnetic flux $\phi/(2\pi) = 1064/769$.
%

%
Using extensive density-matrix renormalization-group method simulations, we mapped out the ground-state phase diagram of the synthetic flux-ladder model as a function of the interaction strength $U$ and the rung hopping $J_\perp$.
For large values of $U$ and $J_\perp$, the model is typically found to be in a Meissner phase.
Moreover, for intermediate values of $U$ and $J_\perp$, the model hosts biased-ladder phases, which are generally stabilized by the presence of rung-wise interactions and can exist on top of superfluids as well as Mott insulators.
%

%
By time-evolving matrix-product states, we studied how the chiral current $\left\langle j_c \right\rangle $ and the leg-population imbalance $\Delta_m$, which are key observables in the Meissner phase and in the biased-ladder phase, respectively, can be probed in the framework of feasible quantum-quench protocols.
In particular, for the Meissner phase, the instantaneous turning on of the leg hopping $J$ in the rung-localized initial state $\ket{R}$ induces a transient chiral current.
Interestingly, it exhibits a similar dependence on the model parameters as the chiral current in the ground state of the corresponding post-quench Hamiltonian.
We showed that this protocol is especially promising for large values of $U$ and $J_\perp$, which is deep in the Meissner phase.
Concentrating on the leg-population imbalance, we showed that an underlying biased-ladder phase leaves signatures in the short-time dynamics that are induced by the instantaneous turning on of the rung hopping $J_\perp$ in a leg-localized initial state $\ket{L}$.
%

%
The results presented in this paper are expected to provide useful guidance to future experimental implementations of flux ladders exploiting synthetic dimensions.

\begin{acknowledgments}
We thank S.~Greschner and L.~Stenzel for helpful discussions.
This work was supported by the Deutsche Forschungsgemeinschaft (DFG,  German  Research  Foundation) under  Project No. 277974659 via Research Unit FOR 2414 and under Germany’s  Excellence Strategy---EXC-2111---No. 390814868.
M.B. and U.S. acknowledge funding through the ExQM graduate school.
L.T. acknowledges additional support from Fundaci\'{o} Privada Cellex, Fundaci\'{o} Mir-Puig, Ministerio de Ciencia, Innovaci\'{o}n y Universidades (Severo Ochoa CEX2019-000910-S, Plan Nacional FIS2017-88334-P, and Ram\'{o}n y Cajal RYC-2015-17890), and Generalitat de Catalunya (SGR1660 and CERCA program).
C.H. acknowledges funding through ERC Grant QUENOCOBA, ERC-2016-ADG (Grant No. 742102).
\end{acknowledgments}
%
%
%
%
%
%
\bibliography{synth_ladder_paper_bib}
%
%
\end{document}